\documentstyle[aps,multicol,epsfig]{revtex}
\newcommand{\av}[1]     {\langle #1 \rangle}

\newcommand{\eqn}[1]    {(\ref{#1})}
\renewcommand{\narrowtext}{\begin{multicols}{2} \global\columnwidth20.5pc}
\renewcommand{\widetext}{\end{multicols} \global\columnwidth42.5pc}

\def\top#1{\vskip #1\begin{picture}(290,80)(80,500)\thinlines \put(
65,500){\line( 1, 0){255}}\put(320,500){\line( 0, 1){
5}}\end{picture}}
\def\bottom#1{\vskip #1\begin{picture}(290,80)(80,500)\thinlines \put(
330,500){\line( 1, 0){255}}\put(330,500){\line( 0, -1){
5}}\end{picture}}

\def\bleq	{\widetext \top{-2.8cm}}
\def\eleq	{\bottom{-2.7cm} \narrowtext}

\def\etal	{{\em et al.}}

\def\bi         {\begin{itemize}}
\def\ei         {\end{itemize}}
\def\beq	{\begin{equation}}
\def\eeq	{\end{equation}}
\def\beqn       {\begin{eqnarray*}}
\def\eeqn       {\end{eqnarray*}}
\def\beqa       {\begin{eqnarray}}
\def\eeqa       {\end{eqnarray}}
\def\f          {\frac}

\def\a          {\alpha}
\def\b          {\beta}
\def\d          {\delta}

\def\et         {\eta}
\def\g          {\gamma}

\def\vph	{\varphi}

\def\bA         {{\bf A}}

\def\bQ         {{\bf Q}}

\def\cF		{{\mathcal{F}}}

\def\vA		{{\vec{A}}}

\def\dag	{\dagger}

\def\mca	{{M_{\mbox{\scriptsize AF}}}}
\def\mis	{{M_{\mbox{\scriptsize SP}}}}
\def\avmca	{{\bar{M}_{\mbox{\scriptsize AF}}}}

\def\aca	{{\a_{\mbox{\scriptsize AF}}}}
\def\ais	{{\a_{\mbox{\scriptsize SP}}}}
\def\bca	{{\b_{\mbox{\scriptsize AF}}}}
\def\bis	{{\b_{\mbox{\scriptsize SP}}}}
\def\bas	{{\b_{\mbox{\scriptsize AF-SP}}}}
\def\ani	{{a_{\mbox{\scriptsize A}}}}
\def\bni	{{b_{\mbox{\scriptsize A}}}}
\def\abr	{{a_{\mbox{\scriptsize B}}}}
\def\bbr	{{b_{\mbox{\scriptsize B}}}}
\def\vphb	{{\vph_{\mbox{\scriptsize B}}}}

\def\vphni	{{\vph_{\mbox{\scriptsize A}}}}
\def\vphb	{{\vph_{\mbox{\scriptsize B}}}}
\def\mni	{{m_{\mbox{\scriptsize A}}}}
\def\mb		{{m_{\mbox{\scriptsize B}}}}

\begin{document}
\draft
\title{Phenomenological Theory of Superconductivity and Magnetism
in Ho$_{1-x}$Dy$_x$Ni$_2$B$_2$C}
\author{Hyeonjin Doh, Manfred Sigrist$^\dag$,
B.K. Cho\footnote{present address :
\it Dept. of Material Science and Technology, K-JIST, Kwangju 500-712 Korea}, 
and Sung-Ik Lee}
\address{National Creative Research Initiative Center for Superconductivity
and Dept.\ of Physics, \\
Pohang University of Science and Technology, Pohang 790-784, Korea\\
$^\dag$Yukawa Institute for Theoretical Physics, Kyoto University,
Kyoto 606-8502, Japan}
\maketitle
\begin{abstract}
The coexistence of the superconductivity and magnetism in the
Ho$_{1-x}$Dy$_x$Ni$_2$B$_2$C is studied by using Ginzburg-Landau theory.
This alloy shows the
coexistence and complex interplay of superconducting and magnetic
order. 
We propose a phenomenological model which includes two magnetic 
and two superconducting order parameters accounting for the
multi-band structure of this material.
We describe phenomenologically the magnetic fluctuations and order and
demonstrate that 
they lead to anomalous behavior of the upper critical field.
The doping dependence of $T_c$ in Ho$_{1-x}$Dy$_x$Ni$_2$B$_2$C
showing a reentrance behavior are analyzed yielding a very good
agreement with experimental data.
\end{abstract}

\begin{multicols}{2}

The discovery of the series of nickelborocarbide,
$R$Ni$_2$B$_2$C ($R$ = Y, Yb, Dy, Ho, Er, Tm, and
Tb)\cite{Nagaragan94,Cava94,Siegrist94}, a few years ago
opened a new chapter in the discussion of magnetic superconductors
\cite{Canfield98}.
This class of materials shows variety of magnetic ordered states\cite{Lynn97} 
depending on the rare-earth atoms $R$, because the magnetism is due to
the spins of their localized $4f$-orbital coupled via RKKY interaction.
Some of these compounds show superconductivity with a comparatively high transition
temperature ($T_c=16.6 K$ for LuNi$_2$B$_2$C).
Band calculations of non-magnetic LuNi$_2$B$_2$C\cite{Mattheiss94,Pickett94}
reveal a complicated multi-band structure with a rather high
density of states at the Fermi level. Superconductivity is likely to
be conventional originating from  electron-phonon interaction due to the
easily polarizable light ions B and C. The relatively large
transition temperature is then a consequence of the enhanced density
of states.

An interesting aspects of this material class is that some 
members ($R$=Tm, Er, Ho, and Dy) exhibit simultaneously
both magnetism and superconductivity.
They have a rather wide range of the ratio of $T_c/T_N$
from $0.6$ for $R$=Dy to $7.0$ for $R$=Tm.
Among them, HoNi$_2$B$_2$C and DyNi$_2$B$_2$C constitute ideal
materials to study the relation between superconductivity and
magnetism. They display the same antiferromagnetic
order, but have different $T_N $ due to a different de Gennes
factor $ (g_J-1)^2J(J+1)$ (dGF). This leads to the situation that the
superconducting $ T_c $ lies above (below) $ T_N $ for the Ho
(Dy)-compound. Therefore, it is natural to 
investigate how the two phases develop in the alloy
Ho$_{1-x}$Dy$_x$Ni$_2$B$_2$C \cite{Cho96prl}.
The experiments show that there is no 
simple de Gennes scaling of $ T_c $ which is expected to monotonically 
decrease with increasing $ (g_J-1)^2J(J+1)$ according to the
Abrikosov-Gor'kov theory, if we assume the R-ions enter as magnetic
impurities. This is not the case here, since 
the R-ions form a regular lattice with magnetic order. In the crossing
region of $ T_N $ and $ T_c $ the superconducting phase boundary as
function of doping concentration $ x $  is apparently
discontinuous as a consequence of a reentrant behavior \cite{Cho96prl}. 
The purpose of this letter is to explain these and further experimental
features consistently within a single phenomenological model including both
magnetism and superconductivity.

The neutron scattering and magnetization experiments of
HoNi$_2$B$_2$C\cite{Canfield94} and DyNi$_2$B$_2$C, show that the
magnetic properties of the both materials are highly anisotropic
with an easy axis in $[110]$-direction
\cite{Lynn97,Grigereit95,Lynn96,Kreyssig97}. 
Both HoNi$_2$B$_2$C and DyNi$_2$B$_2$C have antiferromagnetic (AF) order
below $ T_N = $ 5 K and 10 K, respectively.
The spins order ferromagnetically within the $a$-$b$ plane and
antiferromagnetically in $c$-direction.
For HoNi$_2$B$_2$C, there exists a significant spiral short-range correlation (SP) 
with the wave vector, $\bQ^* \simeq 0.91\bQ$, 
where $\bQ$ is the AF wave vector, i.e. 
the spins are aligned ferromagnetically in the $a$-$b$ plane,
and rotate by $\phi\sim 163^\circ$ between adjacent planes.
A plausible microscopic origin of these two magnetic correlations has 
been discussed by Kalatsky et al. \cite{Kalatsky98} and 
Amici et al. \cite{Amici98}. 

Based on this experimental information, we construct our phenomenological
model of two magnetic order parameters, $ M_{\rm AF} $ and $ M_{\rm SP} $,
corresponding to the two dominant correlations. The effective 
Ginzburg-Landau free energy has the form, 
\bleq
\beq
\cF_{M}=\aca(T)\mca^2 + \ais(T)\mis^2+
\f{\bca}{2}\mca^4+\f{\bis}{2}\mis^4
+ \bas\mca^2\mis^2,
\eeq
\eleq
\noindent
where $\aca(T)=\aca_0(T-T_{\rm AF})$ and $\ais(T)=\ais_0(T-T_{\rm SP})$.
For the sake of simplicity we restrict to the two dominant wave 
vectors ($ {\bf Q} $ and $ {\bf Q}^* $) of the spatial fluctuation
only and take their mode-mode coupling into account \cite{Moriya85}. 
We assume that the AF order is dominant ($ T_{\rm AF} > T_{\rm SP} $)
and finally develops long-range 
order. Once the AF order is established, the SP order
is suppressed, since the two magnetic orders compete with each other
described by the mode-mode coupling term ($ \beta_{AF-SP} > 0 $).  
Thus, only the AF order parameter has a non-zero mean
value, as the temperature is lowered. Considering the fluctuation
effects, we separate the order parameters into mean
value and fluctuation part,
\beq
\mca = \avmca + \d\mca \quad {\rm and} \quad \mis = \bar{M}_{\rm SP} +
\d \mis 
\eeq
where $ \bar{M}_{\rm SP} = 0 $ for all temperatures. 
The mean value of the AF order parameter is determined
by minimizing the free energy with respect to $\avmca$, including the
renormalization due to fluctuations,
\beq
\avmca^2 = \f{\aca(T)+3\bca \av{\d\mca^2} + \bas \av{\d \mis^2} }{\bca}
\label{meanM}
\eeq
The N\`eel temperature $ T_N $ is defined by the vanishing of $
\avmca^2 $. In the calculation of the mean square of the fluctuation in the
transition region the renormalization due to the presence of the  
fourth-order (mode-mode coupling) terms has to be included. This can
be done approximately by applying a standard self-consistent decoupling 
\beqa
\d\mca^2&=&\av{\d\mca^2}+(\d\mca^2-\av{\d\mca^2}),
\label{mf1} \\
\d\mis^2 &=& \av{\d\mis^2}+(\d\mis^2-\av{\d\mis^2}),
\label{mf2}
\eeqa
leading to effective Gaussian fluctuation model \cite{Moriya85}. Thus,
substituting Eq.(\ref{mf1}) and (\ref{mf2}) into the free energy 
we obtain the following expression
up to second order in the fluctuations of the order parameter,
\bleq
\beqa
\d\cF_{M}
&=&\left(\aca(T)+3\bca\avmca^2+\bca\av{\d\mca^2}
+\bas\av{\d\mis^2}\right)\d\mca^2-\f{\bca}{2}\av{\d\mca^2}^2\\
&&+\left(\ais(T)+\bas\avmca^2+\bis\av{\d\mis^2}
+\bas\av{\d\mca^2}\right)\d \mis^2-\f{\bis}{2}\av{\d \mis^2}^2  \nonumber 
\eeqa
\eleq
\noindent
From this we can calculate self-consistently the Gaussian fluctuations 
including Eq.\eqn{meanM} \cite{Moriya85}
\beqa
\label{fluct_eq1}
\av{\d\mca^2}&=&\int\!\!d\d\mca~\d\mca^2 e^{-\b \d\cF}/
\int\!\!d\d\mca e^{-\b \d\cF}\\
\av{\d \mis^2}&=&\int\!\!d \d \mis~\d\mis^2 e^{-\b
\d\cF}/\int\!\!d \d \mis~e^{-\b \d\cF} 
\label{fluct_eq2}
\eeqa

\begin{figure}[bh]
\vspace{-5mm}
\centerline{\epsfxsize=8cm \epsffile{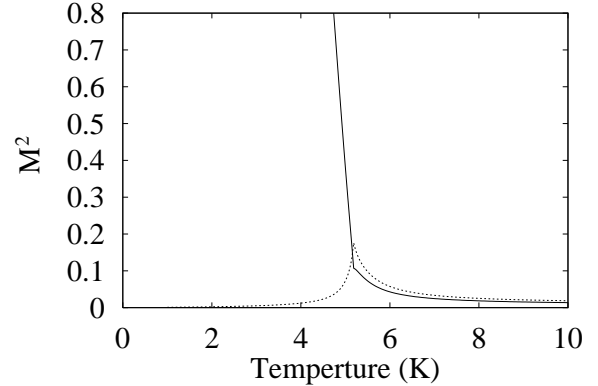}}
\caption{Intensity of the antiferromagnetic $\mca$ and $c$-axis
spiral $\mis$. The solid line denotes the intensity of 
the antiferromagnetic
order and the dotted line the intensity 
 of the spiral order.\label{fig1}}
\vspace{-6mm}
\end{figure}

Since the modulation of  the magnetization in the Ho$_{1-x}$Dy$_x$Ni$_2$B$_2$C
occurs with a short length scale, i.e. has large wave vector
$ {\bf Q} $ or $ {\bf Q}^* $, the superconductivity does not
respond through coupling to the magnetic field (vector potential). 
For an (spin singlet) s-wave superconductor the local spin polarization 
is pair breaking such that the above magnetic order should repel
superconductivity for electrons in orbitals subject to a net spin
moment due to the  
$4f$-spin ordering. (A quantitative estimate of this effect
requires rather detailed knowledge about the complicated band
structure. Interestingly in experiment the N\'{e}el
temperature does not 
deviate much from the de Gennes scaling, i.e. it seems to be basically 
not affected by the superconducting order. 
Thus, we neglect the effects of the superconductivity on the magnetism.)
Now we can calculate $ \av{M_{\rm AF}^2 }$ and $ \av{M_{\rm SP}^2}
$ using the above self-consistent scheme.
The result of the temperature dependence of the intensities $ \av{M^2}
$ is shown
in Fig.\ref{fig1}.  We choose the parameters to
obtain a qualitative agreement with the neutron scattering
experiments of the pure Ho-compound \cite{Lynn97}. This yields a bare
transition temperatures $ T_{\rm AF} $ and $ T_{\rm SP} $ which differ only by 
a few percent.
Here the SP phase grows as temperature decreases and
disappears quickly below the onset of AF order.

According to the band calculation of LuNi$_2$B$_2$C\cite{Mattheiss94},
the Ni(3d) band has the largest contribution to the density of states
at the Fermi surface suggesting that it dominates in the formation of
the superconducting state. However, also the other bands involving
hybridizations with B($2p$), C($2p$) and Ho($5d$) or Dy($5d$) should
be included in the superconductivity.
Hence, superconductivity of the
nickelborocarbide materials cannot be described by a single-band model. 
In connection with the magnetic order there is clear
distinction among the bands depending on the location of the 
each element within the unit cell. For the
AF order the influence of the rare-earth magnetic moments
is canceled on the Ni-sites located exactly in the
center of a tetrahedron of the nearest Ho (Dy) atoms.  
The band originating from Ni(3d) does not feel the 
magnetic momentum of Ho(Dy) anymore below the N\'{e}el temperature.
On the other hand, the field generated by the moments
of Ho(Dy) is not canceled at the site of the other elements.
Moreover, because of large in-plane ferromagnetism of Ho(Dy) 
magnetic moments, the other ions feel a comparatively strong (pair
breaking) spin polarization.

In the case of SP correlation, the magnetic field at the Ni site is not 
canceled exactly. Thus, the SP phase would also affect
the superconductivity of Ni(3d) band.
For other ions the effect of the SP correlation is not so important
since the dominant effect comes eventually from AF order.
Hence we propose simple phenomenological model which consists of two
superconducting order parameters. 
One superconducting order parameter $ \varphi_A $ is associated with Ni(3d) band, and
couples only to the SP order parameter. The other ($ \varphi_B $) 
is due to the other bands involving essentially all elements. This
superconductivity is suppressed by the AF order. 
The Ginzburg-Landau free energy for the these two superconducting
order parameter is
\bleq
\beqa
\label{gleqn_sc}
\cF_{\mbox{\scriptsize SC}} &=& 
\int\!\!d^3r~~\left[ \f{\hbar^2}{2\mni}|(\nabla-\f{2\pi i}{\phi_0}\vA)\vphni|^2
+\ani (T)|\vphni|^2 + \f{1}{2}\bni|\vphni|^4 
+\f{\hbar^2}{2\mb}|(\nabla-\f{2\pi i}{\phi_0}\vA)\vphb|^2
+ \abr (T)|\vphb|^2 + \f{1}{2}\bbr|\vphb|^4 \right. \nonumber \\
&& \left. -\g_1(\vphni^\ast\vphb + \mbox{c.c})
-\g_2\left[(\nabla+\f{2\pi i}{\phi_0}\vA)\vphni^\ast
(\nabla-\f{2\pi i}{\phi_0}\vA)\vphb  + \mbox{c.c}\right]
+ \et_1 \mca^2|\vphb|^2+\et_2 \mis^2|\vphni|^2  \right] ,
\eeqa
\eleq
\noindent
where the parameters are chosen as $T_{cA}=6.2$K, and $T_{cB}=2.9$K ($
\phi_0$ : standard flux quantum). 

For the superconductivity, we obtain two coupled differential equations
for superconducting order parameters from the Ginzburg-Landau free energy
\eqn{gleqn_sc}. 
If we assume Landau gauge, $\bA=(0,H x,0)$, then the linearization
leads to the following homogeneous equations after solving the
differential equations,
\bleq
\beqa
0&=&\left(\f{\hbar^2}{2\mni}\f{2\pi H}{\phi_0}
+\ani(T)+\et_2 \av{\mis^2}\right)\vphni 
- \left(\g_1 -\g_2\f{2\pi H}{\phi_0}\right)\vphb \\
0&=&\left(\f{\hbar^2}{2\mb}\f{2\pi  H}{\phi_0}
+\abr(T) +\et_1 \av{\mca^2}\right)\vphb 
- \left(\g_1 -\g_2\f{2\pi H}{\phi_0}\right)\vphni
\eeqa
\eleq
\noindent
The superconducting instability corresponds to the vanishing of the
determinant of these two equations which determines $T_c$ and
$H_{c2}$. Note that in the absence of an external field the onset of
superconductivity occurs at $ T_c $ higher than both $ T_{cA} $ and $
T_{cB} $ as a result of the coupling of the order parameter components.

Now we include the Dy-doping in HoNi$_2$B$_2$C.
Dy and Ho has same magnetic ordering properties and similar magnetic moments
except for the magnitude of 
its dGF which determines the strength of
the coupling between localized $4f$-spin and conduction electrons.
Thus, we assume the doping of Dy changes only the average dGF which
grows linear with the Dy concentration $x$ ($ {\rm dGF}(x) = {\rm
dGF(Ho)} \times (1-x) + {\rm dGF(Dy)} \times x $).
We take $\et_1$ and $\et_2$ to be linearly proportional 
to the dGF. Note, that the N\'{e}el temperature due to RKKY interaction
depends also linearly on the average dGF.

\begin{figure}[ht]
\vspace{-4mm}
\centerline{\epsfxsize=9cm \epsffile{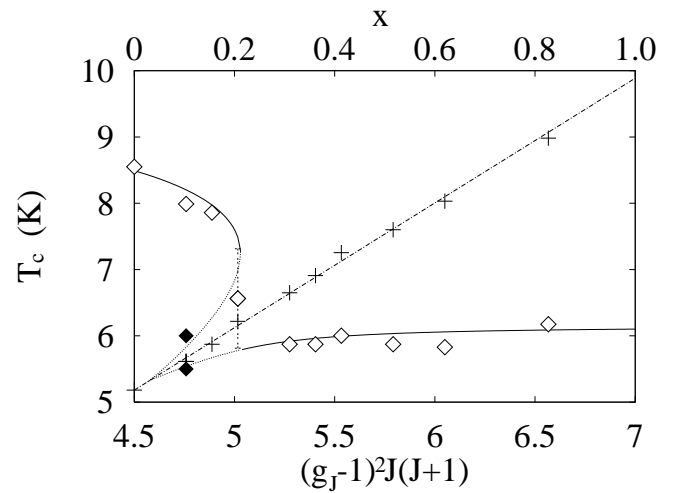}}
\caption{dGF and doping concentration vs. $T_c$
for Ho$_{1-x}$Dy$_x$Ni$_2$B$_2$C. 
The dashed line denotes the N\'{e}el temperature and
the solid line the $T_c$. The dot-dashed line indicates the phase
boundary in the reentrant region.
The diamond points mark the $T_c$, 
the plus points the $T_N$, and the filled diamond points
the boundary of the reentrant region in the experiments [11]. \label{fig2}}
\vspace{-5mm}
\end{figure}

As shown in Fig.\ref{fig2}, our calculation explains well the
experimental $T_c$ variation even when $T_c<T_N$. 
At small doping ($x<0.2$), the $T_c$ determined by
both the $\vphni$ and $\vphb$,
decreases due to the AF and SP fluctuations which also
introduce an apparent discontinuity of the
onset of superconductivity because of a reentrant normal state region near
the $T_N$. For $ x > 0.2 $ the upper superconducting
region ceases to exist. The lower $T_c$ increases slightly with
growing $ T_N $ and remains as the only
superconducting transition for larger Dy-doping concentrations. 

\begin{figure}[ht]
\vspace{-4mm}
\centerline{\epsfxsize=8cm \epsffile{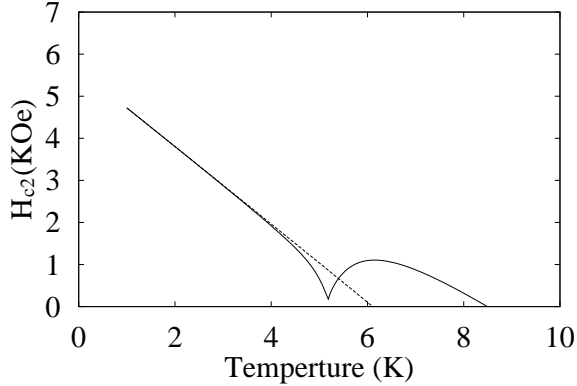}}
\caption{Temperature vs. $H_{c2}$. The
solid line denotes the $H_{c2}$ of HoNi$_2$B$_2$C, and
the dotted line denotes the $H_{c2}$ of DyNi$_2$B$_2$C.\label{fig3}}
\vspace{-4mm}
\end{figure}

Our two component order parameter model gains further support, if we compare
the $H_{c2}$ curves of HoNi$_2$B$_2$C\cite{Rathnayaka96} 
and DyNi$_2$B$_2$C\cite{Cho95}.
Fig.\ref{fig3} shows the $H_{c2}$ of our model for 
HoNi$_2$B$_2$C and DyNi$_2$B$_2$C. 
Our result reproduces the dip 
in the upper critical field of Ho-compound
near the N\'{e}el temperature as in Ref.\cite{Rathnayaka96}.
Below the N\`{e}el temperature, both  $H_{c2}$ curves are 
more or less identical, because the remaining superconducting (Ni)
band feels basically the same magnetic environment in both the Ho- and Dy-compound. 

As our model describes some of the physical properties of the
alloy Ho$_{1-x}$Dy$_x$Ni$_2$B$_2$C very well, we may ask whether some
conclusions can be drawn also for related systems. Starting from
DyNi$_2$B$_2$C, the Dy-ions may be replaced by other rare-earth
elements. Since the superconductivity in this compound relies 
on the subtle canceling of the ordered magnetic moments on the
Ni-site, other elements (different from Ho or Dy) disturb this
balance substantially.
Thus, doping DyNi$_2$B$_2$C with non-magnetic elements, Lu,
(Lu$_{x}$Dy$_{1-x}$Ni$_2$B$_2$C) yields a large net moment on the Ni-site
such that Lu should act like a magnetic impurity. Indeed the $ T_c $
decreases with the doping concentration of Lu in agreement with the
Abrikosov-Gor'kov theory. In addition, $ T_N $ also drops due to
dilution of the Dy-moments. A similar behavior is expected, if we dope
the Dy-compound with
a magnetic element whose crystal field and spin-orbit
coupling effects support a different ordering. Also such dopants
disturb the balance and act as a magnetic impurity on the Ni-based
superconductivity. Furthermore, these elements introduce
disorder in the 4f-spin system such that simultaneously $ T_N $ also decreases
with doping concentration. 
The above alloy gives results consistent with our
assumption that the band relevant for superconductivity in the Dy-rich 
compound is Ni-based.

From our model, which describes the behavior and mutual influence of
two magnetic and two superconducting order parameters, we can derive
a consistent theory explaining all the key experiments in the alloy
Ho$_{1-x}$Dy$_x$Ni$_2$B$_2$C. The multi-band structure is a crucial
aspect to understand the superconducting properties in this system.
It is an important feature that the Ni($3d$) which dominates the
superconductivity does not couple to the spin ordering of the rare
earth ions, Ho and Dy. Thus, the mutual interaction of magnetic order and
superconductivity is mainly due to the other bands which seem to be
weaker superconductors readily suppressed by the magnetism. 
We have found that our model develops a 
reentrance behavior near the crossing region of $ T_c $ and $ T_N $. 
This reentrance can also explain qualitatively 
the non-monotonic temperature
dependence of the electrical resistance for compounds in this doping
region. It is basically a consequence of a sequence of
normal-superconducting-normal-superconducting phase as temperature is lowered. 
Although the reentrant behavior was also reported in polycrystalline
HoNi$_2$B$_2$C, our theory verifies that it appears without any inhomogeneity.
Thus the reentrant behavior of
Ho$_{1-x}$Dy$_x$Ni$_2$B$_2$C is an intrinsic effect.
This region may be very interesting to investigate various fluctuation
effects in and close to the superconducting phase. Clearly disorder
effects must play an important role here too. Thus experimental studies
make only sense with high quality homogeneous samples.
In conclusion, we would like to emphasize that despite the fact that
the superconducting phase is probably electron-phonon induced and
conventional s-wave type, the physics resulting from the interplay
with magnetism and the multi-band effects yield a wealth of unusual 
properties reviving again the study of magnetic superconductors.

H.D. would like to thank the Yukawa Institute for Theoretical Physics
for hospitality during the period of this study.
This work is supported by Creative Research Initiatives
of the Korean Ministry of Science and Technology and by a Grant-in-Aid
of the Japanese Ministry of Science, Education, Culture and Sports.

\newpage
\end{multicols}
\end{document}